\documentclass[a4paper,10pt]{article}
\usepackage[utf8]{inputenc}
\usepackage{multirow} 
\usepackage{booktabs} 
\usepackage{amssymb}
\usepackage{amsmath}
\usepackage{setspace}
\usepackage{graphicx}
\usepackage{amsmath}
\usepackage{color}
\usepackage{gensymb}
\usepackage[colorlinks,citecolor=blue,urlcolor=blue,linkcolor=blue]{hyperref}
\usepackage{authblk}
\title{Sensitivity analysis on imaging the calcaneus using microwaves}
\author[1]{Jesús E. Fajardo}
\author[2]{Fernando Vericat}
\author[3]{Guadalupe Irastorza}
\author[4]{Carlos M. Carlevaro}
\author[5]{Ramiro M. Irastorza}
\affil[1,2,4,5]{\textit{Instituto de F\'isica de L\'iquidos y Sistemas Biol\'ogicos CONICET - CCT La Plata.}\newline \textit{La Plata, Buenos Aires, Argentina}}
\affil[3]{\textit{Centro Diagn\'ostico Mon.}\newline \textit{La Plata, Buenos Aires, Argentina}}
\affil[4]{\textit{UDB F\'isica, FRBA, Universidad Tecnol\'ogica Nacional.}\newline \textit{CABA, Argentina}}
\affil[5]{\textit{Instituto de Ingenier\'ia y Agronom\'ia, UNAJ.}\newline \textit{Florencio Varela, Buenos Aires, Argentina}}

\begin{document}

\maketitle

\begin{abstract}

The bone quality is asociated with changes in its dielectric properties (permittivity and conductivity). The feasibility of detecting
changes in these properties is evaluated using a tomographic array of 16 monopole antennas with z-polarized microwaves
at 1.3GHz. The direct problem was evaluated computationally with the Finite-Difference-Time-Domain (FDTD) method. Local and global sensitivity analysis were considered for identifiyng the parameters that most affect the detection. We observed that the direct problem is highly sensitive to the conductivity of the tissues that surround the calcaneus and the one of the calcaneus itself. Global and local sensitivity methods have shown evidences for feasible detection of variation in dielectric properties of bone. 
\\

\end{abstract}

\section{Introduction}
Microwave imaging is a consolidated research topic on sensing objects non-invasively in several areas; including geophysical science, industrial, and biomedical engineering \cite{pastorino2010microwave}. This technique is well suited for imaging biological tissues due to its non-ionizing condition. In biomedical microwave imaging there are two emerging clinical applications: breast tumour detection \cite{meaney2007initial}, and stroke diagnosis \cite{persson2014microwave}. Particularly, microwave tomography (MWT) is dedicated to estimate the dielectric properties (relative permittivity ($\varepsilon_{r}$) and conductivity ($\sigma$)) of the scatterer object interrogated by an array of antennas.

Recently, efforts were done in order to imaging bone tissue of extremities \cite{semenov2007microwave,meaney2012clinical,gilmore2013microwave}. Meaney et al. \cite{meaney2012clinical} have obtained the first image of calcaneus \textit{in vivo} using MWT, they have compared two patients who had one leg injured and required reduced weight bearing during the healing process (consequently, with a diminished bone mass). The authors have measured different dielectric properties of injured heels as compared to the healthy of each patient. The calcaneus is composed of a body of trabecular bone surrounded by a thin layer of cortical bone. Its microstructure and cortical thickness change depending on many factors, among others, the age, the physical activity, etc. \cite{sornay2007alterations,boutroy2005vivo}. The microstructure properties are related to the bone health, for example, an osteoporotic bone has a higher porosity than a healthy one. It has been measured that a change in bone volume / total volume ratio (BV / TV) produces variations in the microwave dielectric properties of trabecular bone \textit{in vitro} \cite{irastorza2014modeling,meaney2012bone}. These findings have revealed that dielectric properties can predict the trabecular bone health.

There are many factors that make \textit{in vivo} bone microwave imaging challenging. For instance, one of the difficulties found in the inversion algorithms is the high contrast of the dielectric properties of low and high water content tissues \cite{semenov2005microwave}. In particular, bone tissue has a significantly lower water content as compared to muscle or skin and this is also reflected in the dielectric properties \cite{gabriel1996dielectric}. It was also found that the thickness of the tissues involved in the measurement may also influences negatively in the estimation of the dielectric properties. In this regard, Gilmore et al. \cite{gilmore2013microwave} have encountered problems on the detection of bone with microwave when the adipose subcutaneous tissue is thick (more than 7 mm). In \cite{nuutinen2004validation} the authors have evaluated local oedema by estimating the water content of subcutaneous fat using microwave dielectric permittivity. This indicates that changes in heel liquid content (particularly notorious in the elderly population) affect the dielectric properties and thickness of the skin and subcutaneous fat, and consequently, the inverse problem as well.  

The first objective of this paper was to quantitatively evaluate how the geometric (thicknesses and bone area) and dielectric properties of the involved tissues affect the direct problem of MWT of the calcaneus. A computational realistic approach of the configuration of antennas used in \cite{meaney2012clinical} was considered. Likely physiological scenarios of changes in cortical and skin thicknesses were simulated. Changes in dielectric properties of the tissues were also simulated. Special attention was paid in the sensitivity of model to variations of the trabecular bone dielectric properties. 

The definition of sensitivity for microwave imaging systems is still under discussion in literature \cite{moussakhani2014sensitivity}. Such a definition could be used as a tool to study the feasibility of applying MWT to a particular zone of the body. In this respect, we proposed both, local and global sensitivity analysis. The definition of the former was inspired in Moussakhani et al. \cite{moussakhani2014sensitivity} and the second analysis was the Morris method \cite{campolongo2007effective}. These methods were based on amplitude-only information of the scattering parameters. The second objective of this work was to select a sensitivity analysis feasible for being used in biomedical MWT. 

\section{Methods}

\subsection{Model geometry} 
\label{modelo}
\begin{figure}[h]
    \centering
    \includegraphics[width=1.0\textwidth]{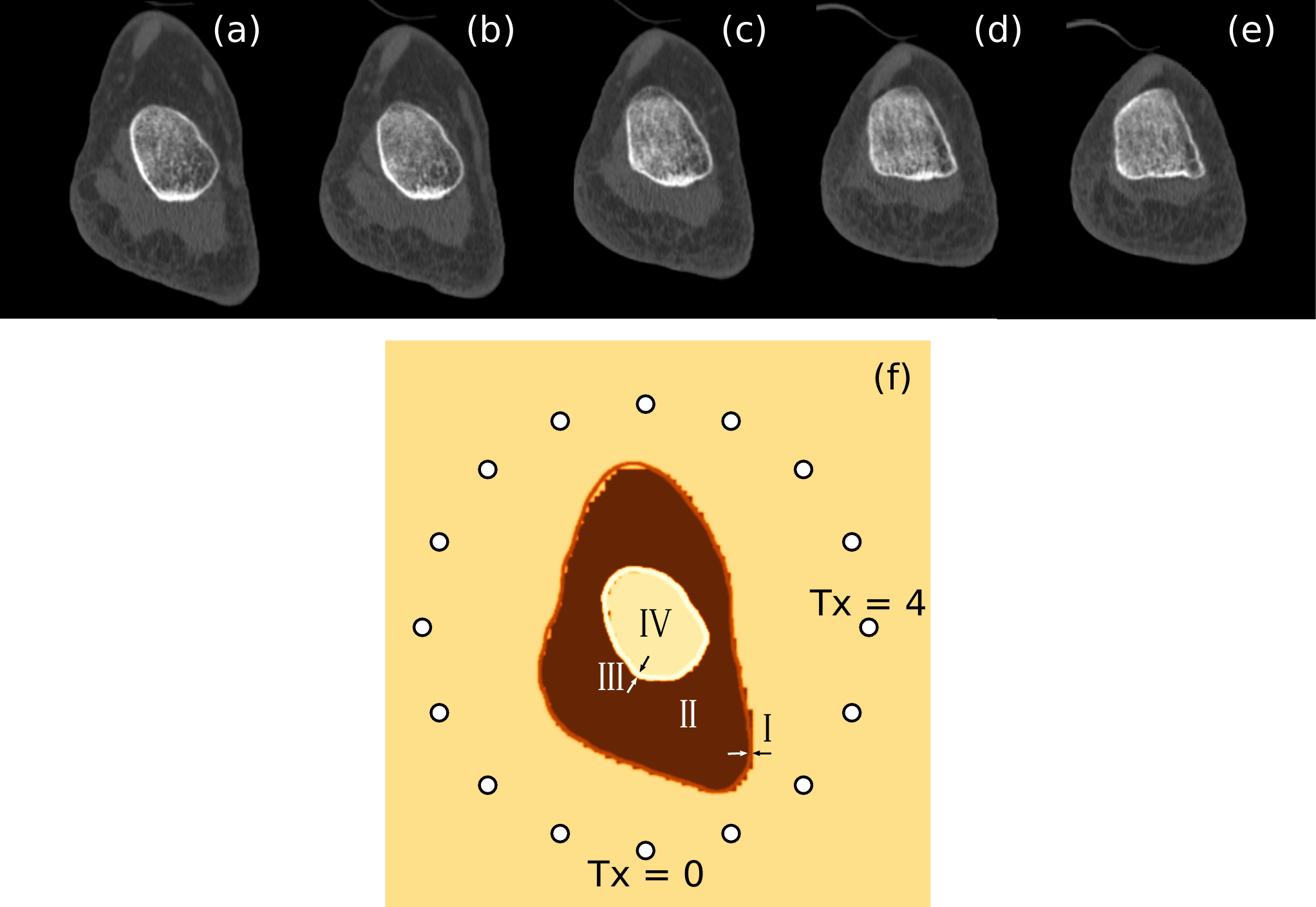}
    \caption{Model geometry. (a-e) The five coronal views (spaced by 4 mm) of the left heel. (f) Regions defined to build the simulation models. White dots represent the circular array of antennas (they are not represented in the real size).}
    \label{Fig1}
\end{figure}

Figure \ref{Fig1} shows the slices from a computed tomography scan of a female left heel. Five coronal views separated each other by 4 mm were used to build the five 2D model geometries. An example of the realistic model based on Fig.\ref{Fig1} (a) is shown in Fig.\ref{Fig1} (f). We have divided the coronal heel geometry into four regions: (I) skin, (II) muscle or tendon, (III) cortical, and (IV) trabecular bone (see Fig.\ref{Fig1} (f)). These regions have different dielectric properties as detailed bellow. The rest of the simulation box is composed by the coupling bath of glycerine and water (80:20). We have chosen the configuration of antennas used in \cite{meaney2012clinical}: a circular array of 16 monopole antennas equally spaced and disposed in a circle with a diameter of 15.2 cm. The monopole antenna is well suited for simulation as it can be modelled as a line of current (a point source in 2D). 

\subsection{Numerical method and boundary conditions}

Maxwell equations were numerically solved using finite-difference time-domain (FDTD) method. It was implemented in the freely available software package MEEP \cite{OskooiRo10}. The model is excited with an electric pulse $E_{\text{z}}$ proportional to $e^{j\omega t}$ (with $\omega = 2\pi f$ where $f$ is the frequency of 1300 MHz) polarized in the z axis trough a transmitter antenna (Tx) and recived by the rest of antennas (Rx). The spatial grid resolution of the simulation box was 0.5 mm, and the Courant factor was 0.5. The boundary conditions were perfectly matched layers (PML).

\subsection{Physical characteristics of the model}
\label{physpar}
If it is not otherwise specified, the revision of the dielectric properties shown in this section will be at a frequency of 1300 MHz, a temperature of 37 $\degree$C, and for in vivo human tissue. Table \ref{Table I} shows the dielectric properties of the tissues involved in the model. As is usual in literature, most of the data were taken from the work of Gabriel et al. \cite{gabriel1996dielectric} (and the references there in), now available on-line and updated by \cite{hasgall2015database}. 
Although the data from reference \cite{gabriel1996dielectric} are accepted by the scientific community, there are several posterior works which measured dielectric properties of tissues \textit{in vivo}. As commented in the subsection \ref{modelo}, the heel is composed by several tissues but we have considered for the model geometry only four regions. We briefly describe the dielectric properties of them from the outer to the deeper region (I to IV). 

The first region is the skin, which is the most \textit{in vivo} characterised tissue. There exist clinical devices for \textit{in vivo} evaluation of its health that use microwave dielectric properties. See reference \cite{nuutinen2004validation}, for example, in which the authors have found that skin subcutaneous fat changes its dielectric properties with the water content (from $\varepsilon_{r} \approx $ 29  to 38, at 300 MHz). The same group have measured the skin as a whole (epidermis, dermis, and deeper layers) \cite{alanen1999penetration} showing values (of permittivity only) around 28. Gabriel C. \cite{gabriel1996compilation} has shown that wet skin (measured in the arms) takes values of 46.8 and 1.08 Sm$^{-1}$ for the permittivity and conductivity, respectively. The most accepted value is that presented in \cite{gabriel1996dielectric} (see Table \ref{Table I}). We have considered these differences by setting a range of the dielectric properties of skin to $\varepsilon_{r} \in$ [28, 46.8] and $\sigma \in$ [1.00, 1.08] Sm$^{-1}$. 

For the second region we have considered two kind of tissues: tendon and muscle. The dielectric properties of tendon \textit{in vivo} are poorly known. The \textit{in vitro} data from bovine presented by \cite{gabriel1996compilation} are $\varepsilon_{r} = $ 46.4, and $\sigma=$ 1.08 Sm$^{-1}$ (similar to that presented in \cite{gabriel1996dielectric}). Regarding muscle tissue, the values fitted by \cite{gabriel1996dielectric} are obtained from different animal models and they are $\varepsilon_{r} = $ 54.3, and $\sigma=$ 1.10 Sm$^{-1}$. Gilmore et al. \cite{gilmore2013microwave} have measured human muscle \textit{in vivo} with values that go from 29.0 to 70.0 and 1.00 to 1.55 Sm$^{-1}$ at 1 GHz, for the permittivity and conductivity, respectively. This intervals cover a range of $\varepsilon_{r} \in$ [29.0, 70.0] and $\sigma \in$ [0.90, 1.55] Sm$^{-1}$. 

The third region is the cortical bone layer surrounding the trabecular bone of the calcaneus. It is difficult to find \textit{in vivo} measurements of this tissue at the frequency range of interest. The only available data from \cite{gabriel1996compilation} are from ovine skull and from human tibia \textit{in vitro} (see Table \ref{Table I}). Therefore we fixed these values for this region.

The trabecular bone tissue is the fourth region and the region of interest of this work. We have measured human trabecular bone \textit{in vitro} \cite{irastorza2014modeling} from elderly population samples at 1200 MHz. We have obtained higher values as compared to that presented in \cite{gabriel1996dielectric} (data from ovine skull). Data from \cite{gabriel1996compilation} were measured at 23 $\degree$C in human samples (\textit{in vitro}) which give values of 18.2 and 0.38 Sm$^{-1}$
for the permittivity and conductivity, respectively. To date the only \textit{in vivo} measurements were performed by Meaney et al. \cite{meaney2012bone}. The values go from $\varepsilon_{r}=$ 12.5 to 16.7 and $\sigma=$ 0.53 to 0.92 Sm$^{-1}$. We have set the range to $\varepsilon_{r} \in$ [12.5, 20.1] and $\sigma \in$ [0.44, 0.92] Sm$^{-1}$. 
\begin{table}
\centering
\caption{Dielectric properties of the involved tissues at 1300 MHz. All tissue data were extracted from \cite{hasgall2015database}, except * that was extracted from \cite{gabriel1996dielectric}, and ** from \cite{meaney2012clinical}.}
\label{Table I}
\begin{tabular}{lccc} 
\hline
Tissue & $\varepsilon_{r}$ & $\sigma$ (Sm$^{-1}$)& Thickness (mm)\\
\hline
Skin wet* & 44.8 & 1.00 & 2.17 \\
Muscle & 54.3 & 1.10 & - \\
Cortical bone & 12.1 & 0.20 & 1.75 \\
Trabecular bone & 20.1 & 0.44 & - \\
Coupling bath** & 23.3 & 1.30 & - \\
\hline
\end{tabular}
\end{table}

We have also varied the thicknesses of the first and third regions. Skin thickness (STh) range of reference \cite{laurent2007echographic} was considered. For the thickness of the third region (cortical layer), we based the simulated range in the references \cite{sornay2007alterations,boutroy2005vivo}, where high-resolution peripheral quantitative tomography technique is used to evaluate architecture of bone tissue. In these references, although no calcaneus measurements are performed, zones with trabecular and cortical bone are evaluated (distal tibia and distal radius). The resulting Cortical thickness (CTh) is within the range of 0.5 mm to 3 mm. The simulated ranges are summarized in Table \ref{Table II}.

\begin{table}
\centering
\caption{Simulated range of the dielectric properties (at 1300 MHz) and thicknesses of the involved tissues.}
\label{Table II}
\begin{tabular}{lccc} 
\hline
Region & $\varepsilon_{r}$ & $\sigma$ (Sm$^{-1}$) & Thickness (mm)\\
\hline
I & [28.0, 46.8] & [1.00, 1.08] & [1.46, 2.88]\\
II & [29.0, 70.0] & [0.90, 1.55] & - \\
III & 12.1 & 0.20 & [0.50, 3.00]\\
IV & [12.5, 20.1] & [0.44, 0.92] & -\\
\hline
\end{tabular}
\end{table}

\subsection{Sensitivity analysis}

The sensitivity of a system is the minimum variation of an input parameter ($x_i$) that results in a detectable output ($y$). In the realistic model presented here, the output signals can change in response to changes in the permittivity and the conductivity of regions I, II, and IV, or to changes in the thickness of the regions I or III. This results in in a total of eight inputs $x_i$. We propose a function $y(\mathbf{x})$ which is computed from measurable magnitudes (S-parameters), and where $\mathbf{x}$ is a vector containing all the inputs of the model. We considered a local and a global sensitivity analysis. The local method is exhaustive and the global is the Morris method. The procedures are detailed next.

\subsubsection{Exhaustive simulations}

In the simulations we have defined the scattering parameter (S-parameter) as the transmission coefficient between antenna $j$ (transmitter, Tx) and $k$ (receptor, Rx), that is, $S^{jk}=E^{k}_{\text{z}}/E^{j}_{\text{z}}$ (where $E^{k}_{\text{z}}$ is the electric field at Rx and $E^{j}_{\text{z}}$ at Tx in z direction). The averaging of absolute values of S-parameters of the receptors whose variations were higher (as compared to coupling media) was computed to account for the whole system. Particularly, in the setup evaluated here, the seven antennas opposite to the transmitter have the highest variations. For instance, for the transmitter Tx = 0 the computed parameter was: 
\begin{equation}
 s_{j = 0} = \frac{1}{7}\sum^{11}_{k=5} \left|S^{0k}\right|
\label{eq2}.
\end{equation}
Therefore we have defined the output ($y$) as: 
\begin{equation}
y = \frac{1}{N}\sum^{N-1}_{j=0} s_{j}
\label{eq3}
\end{equation}
where $N$ is the number of considered transmitter antennas. For instance, if transmitter 0 is used only, then $y=s_{0}$.

We have performed a set of exhaustive simulations, as a local sensitivity method, and computed $y$ parameter for each of them. For regions I and III we have simulated three values equally spaced of STh and CTh, and combined them in 5 different models. We have applied the same reasoning for $\varepsilon_r$ and $\sigma$ of the region I. For regions II and IV, five values of $\varepsilon_r$ and $\sigma$ were simulated, which combined make 9 simulations for each region. This was computed for $j=0,\ldots ,15$ (transmitters) which makes 448 simulations for each coronal view, and results in a total of 5$\times$448 = 2240 simulations. 

For processing these exhaustive simulations we have also computed the slope of the output $y$ when it is plotted for each $x_{i}$. For example, to evaluate the effects due to a change in the conductivity of region IV, $\sigma_{IV}$ is varied in the range described in subsection \ref{physpar} remaining constant the other parameters (setting their values to that shown in Table \ref{Table I}). Then $\partial s_{j}/\partial \sigma_{IV}$ is estimated by fitting a linear relationship. This gives an estimation of a sensitivity of the model.

\subsubsection{Morris method}
In order to ranking the degree of influence of the input parameters in our model, Morris sensitivity analysis was carried out \cite{morris1991factorial}. This method was chosen due to its computational efficiency as compared to other global sensitivity methods (lower number of evaluations of the model). Regardless it is considered rather qualitative than quantitative, it does not require a high sampling interval in each input variable, 
which suits well enough this work purposes.%, because of the biological variability and irregular shapes of the used samples.\\

In the Morris method, the space $\Omega$ of input variables $x_i$ constitutes an hypercube. The length of each side of the hypercube is 1, since the range of variation of each  $x_i$ 
is scaled to the [0,1] interval. The variables are discretized such that each point of the grid in the i-th direction is multiple of $1/(p-1)$, being $p$ the number of levels at which is divided each variable.\\
The algorithm operates relying in the concept of the elementary effect ($d_{i}$), which is done in the one-factor-at-a-time sense. This means that a set of input parameters in $\Omega$ is selected and its output $y(\mathbf{x})$ is computed, then the the i-th variable is varied by $\pm\Delta$ and the output is computed again. This single step results in a new set of parameters ($\mathbf{x'}$) and its output $y(\mathbf{x'})$. From here, we can calculate the so-called elementary effect as:
\begin{equation}
d_{i}=\frac{\vert y(x_{1}, \cdots, x_{i}\pm\Delta,\cdots,x_{8})-y(\mathbf{x})\vert}{\Delta}
\end{equation}
A total of $p+1$ evaluations are made in each direction through the hypercube. Random paths in the mesh of $\Omega$ are generated to compute the local main effects and higher order interactions between input variables, which makes the Morris method a method for global sensitivity analysis. Thereafter, two values are computed: $\mu^{*}$ (measure of central tendency of $d_{i}$) and $\text{std}$ (standard deviation of $d_{i}$, this parameter account for a non-linear effect on the output or an input involved in interaction with other factors) \cite{saltelli2004sensitivity}.\\
In this work, the evaluation of $y(\mathbf{x})$ was done with the Eq. \ref{eq3}, particularly with the transmitter antennas 0 ($y = s_{0}$) and 4 ($y = s_{4}$) in the first and last coronal slice. In this way, four Morris analysis were performed. The total numbers of input variables was 8 and we selected a total of 10 samples of each $x_{i}$, which results in 10$\times$(8+1) = 90 simulations for each transmitter and a total of 360 simulations. 

The Morris method was implemented computationally via Python programming language, using the SALib library \cite{Herman2017} which provides the Morris method with the improvements proposed by Campolongo et al. \cite{campolongo2007effective}.

\section{Results and Discussions}
The ratio bone area / total area (BA/TA) of all slices was computed in order to compare them and to evaluate which position is better for a potential 2D application. The obtained values from slice (a) to (e) (Fig.\ref{Fig1}) were: 0.162, 0.179, 0.189, 0.219, and 0.237, respectively. The calcaneus area remains almost constant around 0.0036 m$^{2}$.

\begin{figure}
    \centering
    \includegraphics[width=1.0\textwidth]{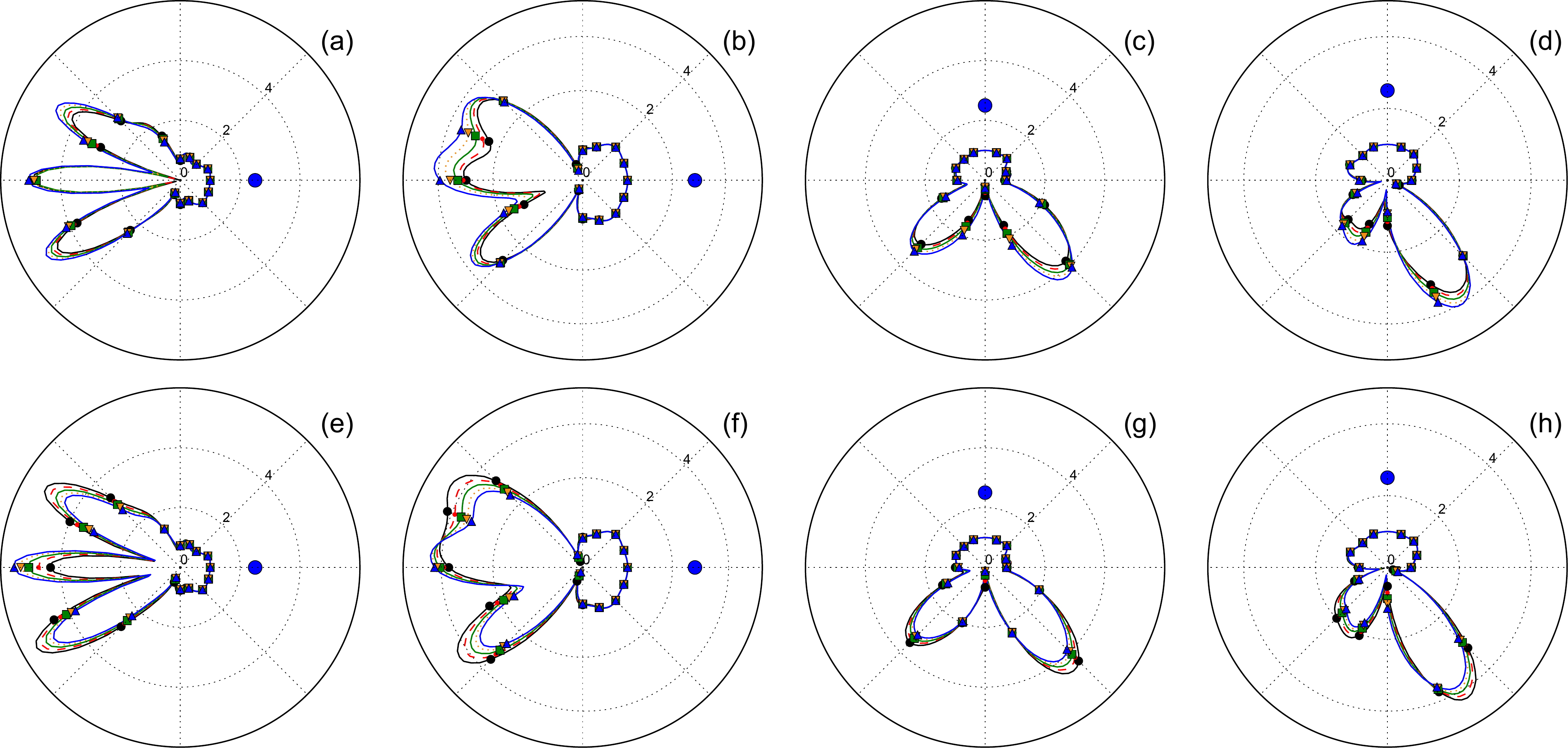}
    \caption{Patterns of normalized transmission coefficients of the array of antennas $\left|S_{s}^{jk} / S_{b}^{jk}\right|$. The lines represent the normalized transmission coefficients in the whole circle where the antennas are disposed, and the smaller symbols (dots,triangles, and squares) the corresponding value of the particular antenna. (a-d) Patterns variating $\varepsilon_r$ of region IV: continuous black, dashed red, continuous green, dotted yellow, and continuous blue lines, correspond to $\varepsilon_r$ = 12.5, 14.4, 16.3, 18.2, and 20.1, respectively. (e-h) Patterns variating $\sigma$ of region IV: continuous black, dashed red, continuous green, dotted yellow, and continuous blue lines, correspond to $\sigma$ = 0.44, 0.56, 0.68, 0.80, and 0.92 Sm$^{-1}$. The position of the transmitter is identified with a blue circle. Columns 1 and 3 corresponds to the coronal slide of Fig.\ref{Fig1} (a). Columns 2 and 4 corresponds to the coronal slide of Fig.\ref{Fig1} (e).}
    \label{Fig2}
\end{figure}

Figure \ref{Fig2} shows the results from varying dielectric properties of the trabecular bone (region IV). The patterns were obtained by computing $\left|\frac{S^{jk}_\text{s}}{S^{jk}_\text{b}}\right|$, where sub-indexes s and b refer to scatterer (with the heel) and background (without the heel, with the matching fluid only), respectively. The results of two very different geometries are shown (slices of Figs.\ref{Fig1} (a) and (e)) together with two transmitters (Tx = 0 and 4). Columns 1 and 3 of Fig.\ref{Fig2} correspond to slice Fig.\ref{Fig1} (a), and 2 and 4 to slice Fig.\ref{Fig1} (e). The main result shown in these figures is that the highest variations were always detected in the seven receptors opposite to the transmitter. The magnitudes of these variations were between two and four times the value of that due to the background, which seems to support the feasibility of the detection by microwave. It also should be noted that the two slices are very different in shape and BA/TA, and it results in very different patterns as well. For instance, for transmitter 0, while column 1 shows three lobes, column 2 shows two lobes. However, when the transmitter was shifted to position 4 two lobes are depicted for both slices. 

\subsection{Exhaustive simulations}
\label{local}
\begin{figure}
    \centering
    \includegraphics[width=0.8\textwidth]{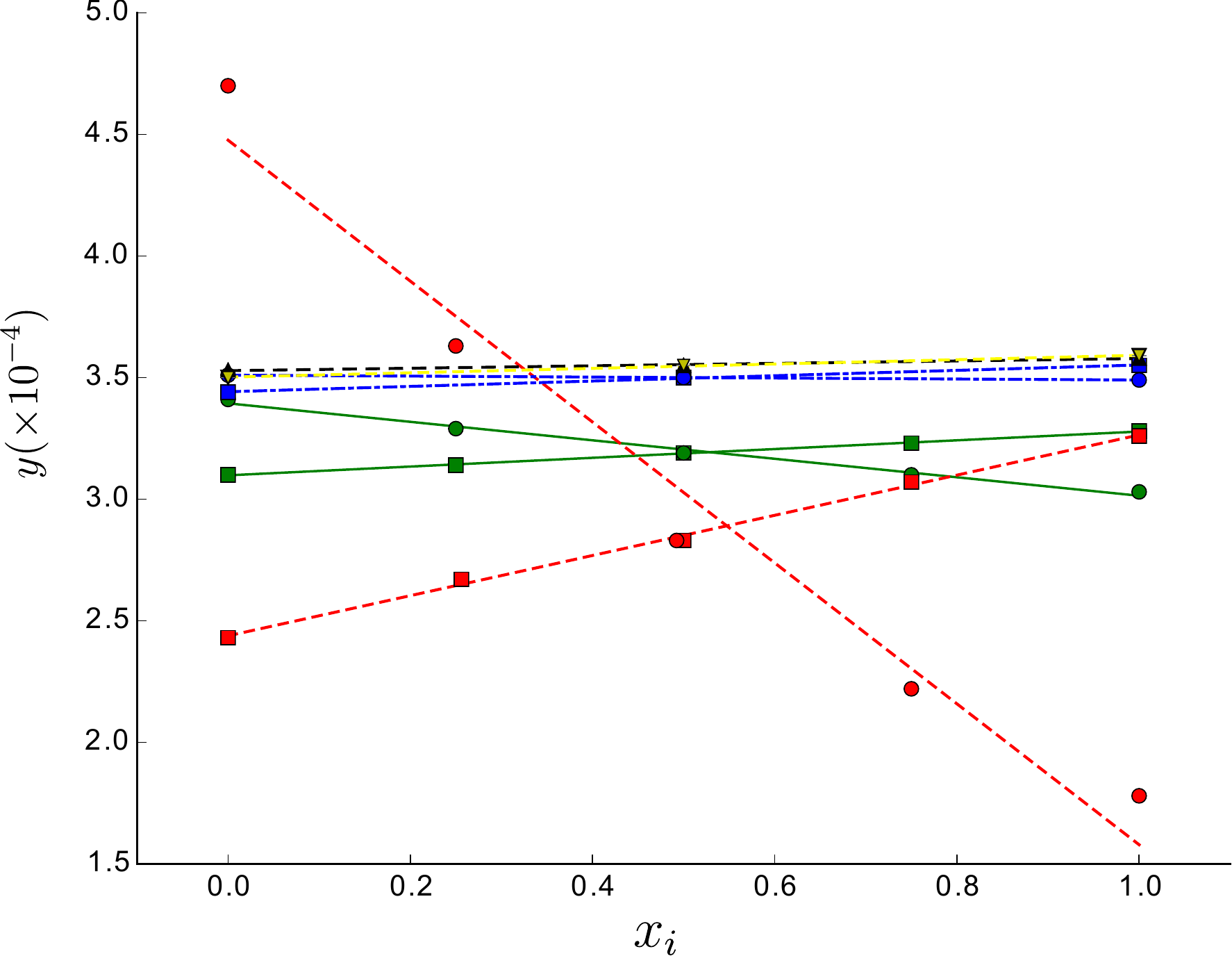}
    \caption{Output $y$ (Eq.\ref{eq3}) with N = 16 for the slice from Fig.\ref{Fig1} (a) (with lowest BA/TA parameter). The blue dashed-dotted, red dashed, and continuous green lines correspond to variation of dielectric properties of region I, II, and IV, respectively. Circles are used for the conductivity ($\sigma$) and the squares for the relative permittivity ($\varepsilon_r$). The yellow and black triangles are used, respectively, for cortical (CTh) and skin (STh) thickness variations. The lines represent the first-order linear fitting used to approximate $\frac{\partial y}{\partial x_i}$.}
\label{Fig3}
\end{figure}

Figure \ref{Fig3} shows an example of $y$ (using Eq.\ref{eq3}) for the slice of Fig.\ref{Fig1} (a) computed with N = 15 (it means using the 16 transmitters). The range of variation of the input parameters $x_i$ (see Table \ref{Table II}) was scaled to the [0,1] interval. Regarding the effect of changing the tissue dielectric properties, these findings qualitatively show that muscle, in particular its conductivity, has the greatest influence in the transmission coefficient. On the other hand, its relative permittivity seems to have higher effects than the trabecular bone relative permittivity and conductivity. The rest of the parameters have considerably lower influence in the output of the model. The lines represented in Fig.\ref{Fig3} are the first-order linear fitting whose slopes were used to approximate $\frac{\partial y}{\partial x_i}$ for each slice.

\begin{figure}
    \centering
    \includegraphics[width=1.0\textwidth]{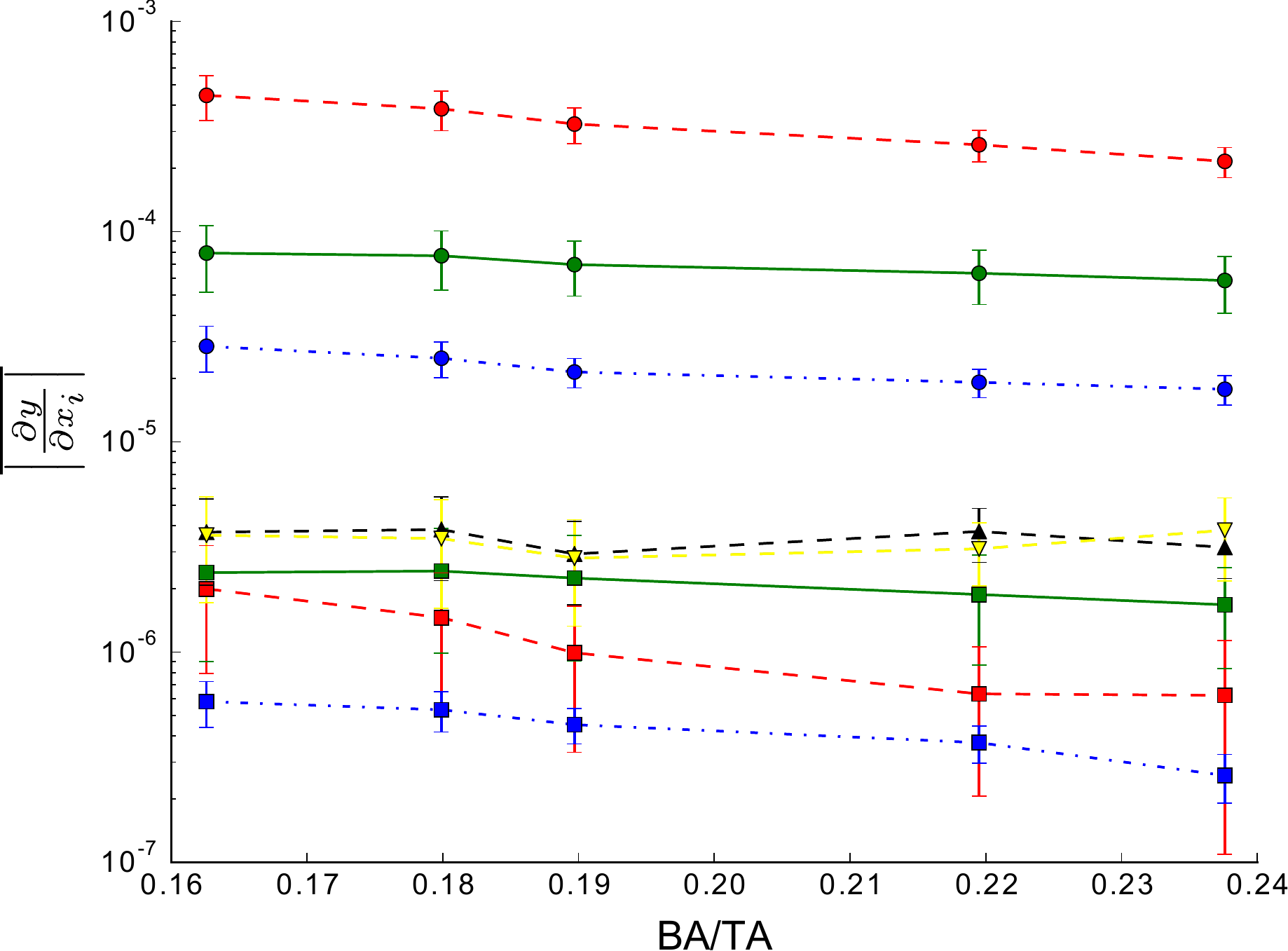}
    \caption{Estimation of local sensitivity as function of the BA/TA ratio. The blue dashed-dotted, red dashed, and continuous green lines correspond to variation of dielectric properties of region I, II, and IV, respectively. Circles are used for the conductivity ($\sigma$) and the squares for the relative permittivity ($\varepsilon_r$). The yellow and black triangles are used, respectively, for cortical (CTh) and skin (STh) thickness variations.}
    \label{Fig4}
\end{figure}

Evaluation of the effects of the ratio BA/TA in the model are shown in Fig.\ref{Fig4}. The mean values of the linear estimation of sensitivity for the eight input parameters are plotted for each slice. The interesting finding is that the greater the BA/TA the lower the absolute value of sensitivity estimation. This evidences that an anterior position of the array of antennas would help the detection of the dielectric properties. Furthermore, the conductivities of tissues were the parameters that have more weight in the sensitivity. Remarkably, the conductivity of region II (muscle or tendon) is followed by conductivities of region IV (trabecular bone) and III (skin), however relative permittivity shows to be a less significant factor in the detection.

\subsection{Global sensitivity}

\begin{figure}
    \centering
    \includegraphics[width=0.8\textwidth]{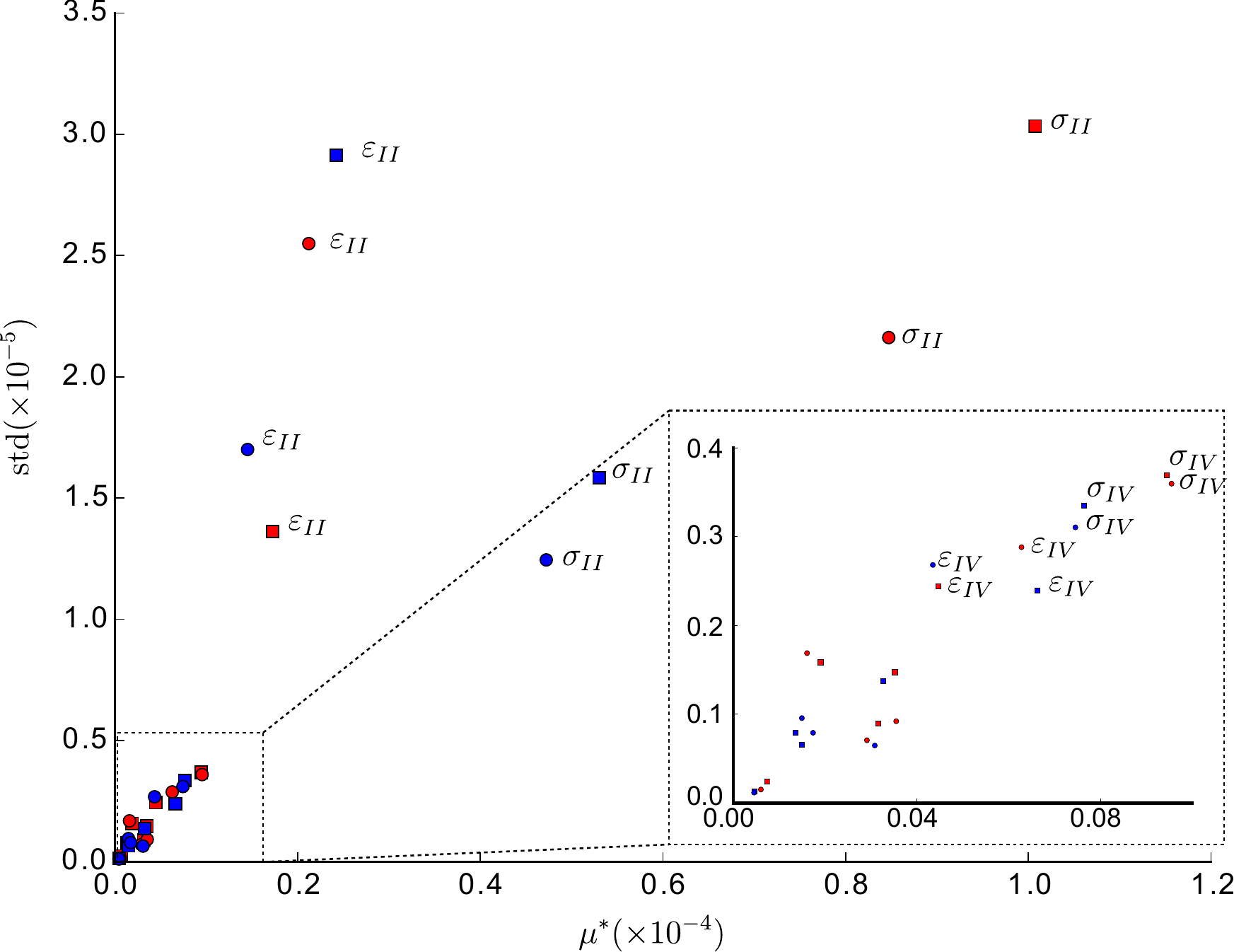}
    \caption{Morris sensitivity results. The results of slices of Fig.\ref{Fig1} (a) and (e) are represented in red and blue, respectively. Circles and squares represent the results using Tx = 0 and Tx = 4, respectively. Inputs with higher sensitivity values are identified. The inset shows the results obtained with the dielectric properties of the trabeculae bone.}
    \label{Fig5}
\end{figure}

In the Morris experiments a large measure of central tendency ($\mu^{*}$) indicates an input with an important influence on the output and a high value of the standard deviation (std) indicates an input with a non-linear effect on the output (or an input involved in interaction with other factors). Figure \ref{Fig5} shows the results of such analysis. For all cases (position of transmitter and slice) the highest values of $\mu^{*}$ were obtained for the conductivity of the region II (muscle or tendon). It means that its effect is quantitatively higher than the conductivity of the trabecular bone, which is in third place (see inset of Fig.\ref{Fig5}). These results are in agreement to the local analysis of subsection \ref{local}. Although relative permittivity of region II produces high values of $\mu^{*}$ (second place), std value is relatively high which means that it has non-linear effects in the output as well. This can explain the low values of $\left|\frac{\partial y}{\partial x_i}\right|$ above obtained (a linear estimate was used). This analysis does not show a significant influence of the rest of the inputs.

As compared to the subsection \ref{local}, it should be remarked that the Morris analysis reduces drastically the number of evaluations of the model and leads to similar results. 

\subsection{Limitation of the study}

This study has certain limitations which are commented next. Firstly, we have focused the objectives of this work in the direct problem, therefore other issues, such as inversion algorithms, were not accounted for. This is outside the scope of this study but need to be assessed to validate the results shown here.

Secondly, we only considered a 2D model with four regions. For region II, tendon could also be considered because, for example, Achilles tendon connects the muscles to the calcaneal tuberosity and it has a well-determined position. The lack of information regarding the dielectric properties of this tissue makes its simulation less reliable. We think that the range covered in the simulations presented here is wide enough, but the region where tendon connects to the calcaneus and the regions occupied by tendons could have different properties from that more distant to it (see the lighter zones in region II of Fig.\ref{Fig1} (a-e)). However, the area covered by this tissue is smaller than that covered by the calcaneus, therefore we do think that results should not change substantially. Regarding experimental validation, other variables should be considered, as for example, the dynamic range and the signal-to-noise ratio of the acquisition equipment. According to the results reported here, the dynamic range needed to measure changes in the dielectric properties of the trabecular bone is $\sim$50dB (calculated from $20 \text{ log} \frac{\overline{S}_{j_{max}}}{\Delta \overline{S}_{j_{min}}}$). The acquisition system recently proposed by \cite{epstein20143d}, has higher dynamic range ($\sim$110dB), which seems to support the feasibility of detecting the dielectric properties of the trabecular bone using microwave tomography.

\section{Conclusions}

In the direct problem of MWT of the heel, 2D models are highly sensitive to the conductivity of the tissues that surround the calcaneus and the one of the calcaneus itself. If the application intends to measure the dielectric properties of the calcaneus, a good estimation of the conductivity of the subcutaneous tissues must be obtained. Morris and a local sensitivity methods have shown evidences for feasible detection of variation in dielectric properties of bone. It was found also that the smaller the ratio of bone area/heel area the greater the sensitivity of detecting the dielectric properties. Finally, the Morris method is a good option for screening this kind of applications since it implies fewer evaluations of the model.

\bibliographystyle{unsrt}
\bibliography{referencias}

\end{document}